\documentclass[%
 reprint,
superscriptaddress,
 amsmath,amssymb,
 aps,
]{revtex4-1}

\usepackage{graphicx}
\usepackage{amsmath}
\usepackage{braket}
\usepackage{url}
\usepackage{dcolumn}

\usepackage[colorlinks,
linkcolor=blue,
anchorcolor=red,
urlcolor=blue,
citecolor=red]{hyperref}

\usepackage[mathlines]{lineno}
\usepackage{changes}
\definechangesauthor[name={SZhang}, color=blue]{szhang}
\definechangesauthor[name={FRXu}, color=red]{frxu}

\begin{document}


\title{{\it Ab initio} descriptions of $A=16$ mirror nuclei with resonance and continuum coupling}
\author{S. Zhang}
\affiliation{
State Key Laboratory of Nuclear Physics and Technology, School of Physics, Peking University, Beijing 100871, China}
\author{F. R. Xu}\email{frxu@pku.edu.cn}
 \affiliation{
State Key Laboratory of Nuclear Physics and Technology, School of Physics, Peking University, Beijing 100871, China}
\affiliation{Southern Center for Nuclear-Science Theory (SCNT), Institute of Modern Physics, Chinese Academy of Sciences, Huizhou 516000, Guangdong Province, China}
\author{J. G. Li}
\affiliation{
Institute of Modern Physics, Chinese Academy of Sciences, Lanzhou 730000, China}
\affiliation{Southern Center for Nuclear-Science Theory (SCNT), Institute of Modern Physics, Chinese Academy of Sciences, Huizhou 516000, Guangdong Province, China}
\affiliation{School of Nuclear Science and Technology, University of Chinese Academy of Sciences, Beijing 100049, China}
\author{B. S. Hu}
\author{Z. H. Cheng}
 \affiliation{
State Key Laboratory of Nuclear Physics and Technology, School of Physics, Peking University, Beijing 100871, China}

\author{N. Michel}
\affiliation{
Institute of Modern Physics, Chinese Academy of Sciences, Lanzhou 730000, China}
\affiliation{School of Nuclear Science and Technology, University of Chinese Academy of Sciences, Beijing 100049, China}
\author{Y. Z. Ma}
\author{Q. Yuan}
 \affiliation{
State Key Laboratory of Nuclear Physics and Technology, School of Physics, Peking University, Beijing 100871, China}
\author{Y. H. Zhang}
\affiliation{
Institute of Modern Physics, Chinese Academy of Sciences, Lanzhou 730000, China}
\affiliation{School of Nuclear Science and Technology, University of Chinese Academy of Sciences, Beijing 100049, China}


\begin{abstract}
We have used an {\it ab initio} Gamow shell model to study the isospin symmetry breaking in the $A=16$ mirror nuclei of $^{16}$F, $^{16}$N, $^{16}$Ne and $^{16}$C. Starting from a chiral interaction with two-nucleon force (2NF) at N$^3$LO and three-nucleon force (3NF) at N$^2$LO, a complex-momentum {\it psd}-shell Hamiltonian was constructed by employing the many-body perturbation theory in the Gamow Hartree-Fock basis which includes bound, resonant and continuum states self-consistently. Such an elaborated {\it ab initio} Gamow shell model with both continuum coupling and 3NF included can properly treat the many-body correlations of weakly bound and unbound nuclei. The mirror partners of $^{16}$F and $^{16}$N exhibit different level orders in their excitation spectra, which can be well explained by the inclusion of 3NF in the calculation. 
The isospin asymmetry between the mirror partners $^{16}$Ne and $^{16}$C was studied in detail by insight into their configuration structures. The interplay between 3NF and the continuum coupling is discussed in the weakly bound and unbound nuclear states. 
\end{abstract}{}

\maketitle
\section{\label{sec:level1}Introduction}

With next-generation Rare Isotope Beam (RIB) facilities, we have the ability to produce most of the rare isotopes located at the edge of the nuclear landscape, thus shedding light on the origin of elements, the fundamental problems of nuclear structure and nuclear forces. However, theoretical descriptions of proton-rich or neutron-rich nuclei in these regions are challenging in terms of theoretical methods and computational demands.
As nuclei approach driplines, the effect of single-particle long-distance asymptotic behavior and the coupling to the continuum are vital in understanding these open quantum systems~\cite{Michel2021}. Indeed, they lead to novel phenomena in weakly bound and unbound nuclei, such as halo ~\cite{TANIHATA2013215}, Borromean~\cite{RevModPhys.76.215,MA2020135673} and Thomas-Ehrmann shift (TES)~\cite{PhysRev.88.1109,PhysRev.81.412}.

Among the open quantum systems, the $A=16$ mirror partners are interesting, and could provide more insights into the isospin symmetry breaking of nuclear force and the evolution of nuclear properties from the valley of stability to driplines. 
The apparent symmetry breaking in the spectra of the mirror partners $^{16}$F and $^{16}$N was observed with significant level inversions starting from the ground states (g.s.), which was explained by the continuum coupling~\cite{PhysRevC.90.014307}. In Ref.~\cite{PhysRevC.102.031303}, energy shifts in pairs of isobaric analog states in mirror nuclei were systematically analyzed. 
It turns out that the large energy splittings of the $^{16}$F-$^{16}$N mirror pair exceed the normal isospin-symmetry-breaking (ISB) behavior, which requires additional ISB effects.  
Theoretically, as mentioned in Ref.~\cite{PhysRevC.90.014307}, the phenomenological Gamow shell model (GSM) and coupled-channel GSM calculations with a dedicated treatment of nuclear correlations have been performed to analyze the $^{16}$F-$^{16}$N mirror pair~\cite{PhysRevC.106.L011301}. It shows that only coupled-channel GSM calculations with corrective factors reproduce the TES observed in the $^{16}$F-$^{16}$N mirror pair. 
For the mirror partners $^{16}$C and $^{16}$Ne, 
larger isospin asymmetries in the configurations of the g.s. and first $2^+$ state are demonstrated and considered as a new mechanism of TES ~\cite{PhysRevLett.88.042502,PhysRevC.91.024325}.
In addition, $^{16}$Ne is an intermediate nucleus of the cascade $2p$ emission of the recently observed four-proton unbound nucleus $^{18}$Mg. The study of mirror asymmetry in the $^{16}$C-$^{16}$Ne mirror pair is meaningful to further understand the nature of higher $2^+$ excitation energy in $^{18}$Mg~\cite{PhysRevLett.127.262502}. Nuclei at driplines exhibit novel phenomena arising from the proximity of the continuum, which also provides a comprehensive and rigorous test of nuclear theory. 

In recent decades, significant progress in {\it ab initio} calculations~\cite{BINDER2014119,Hagen2016,PhysRevLett.117.172501,PhysRevLett.120.152503,Gysbers2019,Hu2022} has been made with the developments of chiral effective field theory ($\chi$EFT)~\cite{RevModPhys.81.1773,MACHLEIDT20111}, similarity renormalization group (SRG)~\cite{PhysRevC.75.061001,BOGNER201094} and many-body methods~\cite{HJORTHJENSEN1995125,CORAGGIO2009135,Hagen_2014,HERGERT2016165,10.3389/fphy.2020.00340}. Meanwhile, three-nucleon force (3NF) has been shown to be crucial in the detailed descriptions of nuclear structure~\cite{PhysRevLett.99.042501,PhysRevLett.105.032501,PhysRevLett.107.072501,PhysRevLett.106.202502,PhysRevLett.108.242501,PhysRevLett.110.242501,PhysRevLett.110.022502,PhysRevLett.113.142501,PhysRevC.98.044305,MA2020135257,PhysRevC.83.031301,PhysRevC.96.014303,PhysRevC.102.054301,PhysRevC.101.014318,PhysRevLett.126.022501,HEBELER20211,Szhang-Chinese}. 
However, as the continuum coupling is essential at the proximity of driplines, weakly bound and unbound nuclei are challenging theoretical studies with standard approaches, such as many-body methods using the HO basis. To include the continuum effect, the continuum shell model~\cite{PhysRevLett.95.042503,PhysRevLett.94.052501} has been developed, taking into account the continuum effect by projecting the model space onto the subspaces of bound and scattering states in a real-energy basis. By introducing an interaction for the coupling to the continuum, the continuum-coupled shell model~\cite{Tsukiyama2015} was suggested to include the continuum effect. Several {\it ab initio} methods, such as the no-core shell model with resonating group method~\cite{PhysRevLett.101.092501,PhysRevC.79.044606}, the single-state harmonic oscillator representation of scattering equations~\cite{PhysRevC.94.064320,Mazur2019}, and the no-core shell model with continuum~\cite{PhysRevLett.110.022505,PhysRevC.87.034326}, have also been successfully applied to open quantum systems of nuclei. 
Another choice for that matter is the Berggren basis~\cite{BERGGREN1968265}, which can treat bound, resonant and continuum states on the same footing. The GSM~\cite{PhysRevLett.89.042501,PhysRevLett.89.042502} is a powerful tool, which provides a full description of the interplay between continuum coupling and inter-nucleon correlations via the use of the Berggren basis and configuration mixing.
{\it Ab initio} methods have also been developed in the frame of the Berggren basis, hence with the continuum coupling included, such as the complex coupled cluster~\cite{HAGEN2007169,PhysRevLett.108.242501}, the complex in-medium similarity renormalization group~\cite{PhysRevC.99.061302}, and the
no-core GSM~\cite{PhysRevC.88.044318,PhysRevC.100.054313,PhysRevC.104.024319}. Added to that, the GSM with a core has been further developed by generating effective interactions from realistic forces~\cite{PhysRevC.73.064307,PhysRevC.80.051301,SUN2017227,HU2020135206}.

The $A=16$ nuclear systems provide remarkable cases of interest, which would lead to new insights into nuclear properties, nuclear forces, and many-body methods due to the emergence of the continuum coupling. In this work, we will thus perform {\it ab initio} GSM calculations for $^{16}$F, $^{16}$N, $^{16}$C and $^{16}$Ne. 
We will depict g.s. energies, excitation spectra, as well as other observables of physical interest. With both 3NF and the continuum coupling considered, we probe the ISB effect in the $A=16$ mirror partners. Our calculations will show the necessity to combine the effects between 3NF and continuum coupling in nuclei close to driplines.

\section{\label{sec:level2} The {\it Ab initio} Gamow shell model with three-nucleon force included}

We employ the intrinsic Hamiltonian of the $A$-nucleon system, 
\begin{equation}
\label{Eq1}
\begin{aligned}
H=&\sum_{i=1}^{A}\left(1-\frac{1}{A}\right) \frac{\boldsymbol{p}_{i}^{2}}{2 m}+\sum_{i<j}^{A}\left(v_{ij}^{\mathrm{NN}}-\frac{\boldsymbol{p}_{i}\cdot\boldsymbol{p}_{j}}{m A}\right)\\
&+\sum_{i<j<k}^{A}v^{\mathrm{3N}}_{ijk},
\end{aligned}
\end{equation}
where $\boldsymbol{p}_i$ is the nucleon momentum in the laboratory coordinate, and $m$ is the nucleon mass, while $v^{\mathrm{NN}}$ and $v^{\mathrm{3N}}$  denote the two-nucleon force (2NF) and 3NF, respectively. In Ref.~\cite{PhysRevC.83.031301}, a chiral 2NF plus 3NF interaction, named EM1.8/2.0, was suggested, in which the 2NF took the next-to-next-to-next-to-leading order ($\mathrm{N^{3}LO}$) of Entem and Machleidt~\cite{PhysRevC.68.041001} and was evolved to a low momentum scale $\lambda_{\text{SRG}}=1.8$ $\text{fm}^{-1}$ by the SRG method. The 3NF was obtained at the next-to-next-to-leading order ($\mathrm{N^{2}LO}$) using a nonlocal regulator with a cutoff of $\Lambda_{3\text{N}}=2.0$ $\text{fm}^{-1}$. The low-energy constants (LECs) $c_{1},c_{3},c_{4}$ appearing in the two-pion-exchange of 3NF have the same values as those in $v^{\mathrm{NN}}$. For the LECs in the one-pion exchange and contact term of the 3NF, the EM1.8/2.0 interaction has $c_{D}=1.264$ and $c_{E}=-0.120$ which were obtained by fitting the $^3$H binding energy and the $^4$He point-charge radius. In many-body calculations, 3NF is usually normal ordered at a two-body level neglecting the residual three-body term~\cite{PhysRevC.76.034302,PhysRevLett.109.052501,PhysRevLett.126.022501,PhysRevC.96.014303,PhysRevC.105.014302}. It has been proved that the EM1.8/2.0 interaction can well reproduces nuclear matter saturation~\cite{PhysRevC.83.031301} and globally reproduce ground-state energies from light to heavy nuclei~\cite{PhysRevC.96.014303,PhysRevLett.126.022501,PhysRevC.105.014302}. We take the same procedures including the same LECs and regulators as in EM1.8/2.0 to construct the chiral 2NF plus 3NF interaction that will be used in the present work. 


The Hartree-Fock (HF) calculation is our starting point. The single-particle HF equation can be written as 
\begin{equation}
\label{Eq2}
\begin{aligned}
&\sum\limits_{\beta}\left [\left(1-\frac{1}{A}\right) \frac{\boldsymbol{p}_{\alpha}^{2}}{2 m}\delta_{\alpha\beta}+\sum\limits_{\gamma\delta}\rho_{\gamma \delta}V^{\mathrm{NN}}_{\alpha \gamma \beta \delta}\right.\\&\left.+\frac{1}{2}\sum\limits_{\gamma \mu \delta \nu}\rho_{\gamma \delta}\rho_{\mu \nu}V^{3\mathrm{N}}_{\alpha \gamma \mu \beta \delta \nu} \right ]\psi(\beta) =e_\alpha\psi(\alpha),
\end{aligned}
\end{equation}
where $\rho_{\gamma \delta} = \sum\limits_{i\leq\epsilon_F}\braket{\gamma|i}\braket{i|\delta}$ is the one-body density matrix, with the sum about $i$ running over all hole states below the HF Fermi surface $\epsilon_F$ of the reference state. 
$V^{\mathrm{NN}}_{\alpha \gamma \beta \delta}$ and $V^{3\mathrm{N}}_{\alpha \gamma \mu \beta \delta \nu}$ stand for antisymmetric 2NF and 3NF matrix elements, respectively. 
The two-body term, $-\frac{\boldsymbol{p}_i·\boldsymbol{p}_j}{mA}$, which is to remove the effect from the center-of-mass (CoM) motion, has been incorporated into $V^{\text{NN}}$. 

To deal with the continuum coupling, we have extended the HF method into the complex-momentum (complex-$k$) plane, which leads to the Gamow Hartree-Fock (GHF) calculation~\cite{ZHANG2022136958,PhysRevC.73.064307,HU2020135206}. The complex-$k$ single-particle GHF equation can be expressed as 
\begin{equation}
\label{Eq3}
\begin{aligned}
&\frac{\hbar^2k^2}{2\mu}\psi_{nlj}(k)+\int_{L^+}dk^{\prime}k^{\prime\,2}U(ljk^{\prime}k)\psi_{nlj}(k^{\prime})\\&=e_{nlj}\psi_{nlj}(k),
\end{aligned}
\end{equation}
where $\mu=m/(1-\frac{1}{A})$, and $k$ ($k^{\prime}$) is defined on a contour $L^+$ in the fourth quadrant of the complex-$k$ plane~\cite{SUN2017227}. $U(ljk^{\prime}k)$ is the complex GHF single-particle potential given by 
\begin{equation}
\begin{aligned}
U(ljk^{\prime}k)=\sum_{\alpha\beta}\braket{k^{\prime}|\alpha}\bra{\alpha}U\ket{\beta}\braket{\beta|k},
\end{aligned}
\end{equation}
where $l$, $j$ are the orbital and total angular momenta of a single-particle orbital, respectively. 
Greek letters denote HO states, so that $\braket{\beta|k}$ is the HO basis wavefunctions $\ket{\beta}$ expressed in the complex-$k$ plane. $\bra{\alpha}U\ket{\beta}$ is the HF single-particle potential which is obtained by solving the real-energy HF Equation~(\ref{Eq2}) in the HO basis~\cite{PhysRevC.99.061302}.
The 3NF enters the GHF solution through $\bra{\alpha}U\ket{\beta}$. In numerical calculations, the GHF equation is solved using the Gauss-Legendre quadrature scheme~\cite{PhysRevC.73.064307} with 35 discretizing points on the contour $L^{+}$, which has been proved to be sufficient to make calculations converged~\cite{HU2020135206,ZHANG2022136958}. The GHF basis is composed of bound, resonant and continuum states. Only the chosen core nucleus which is at a closed shell is performed with the HF and GHF calculations to generate the Berggren basis for many-body GSM calculations of nuclei located in a valence space on top of the closed core. 

In {\it ab initio} many-body calculations, the full inclusion of 3NF is computationally costly and often renders calculations impossible. Therefore, the normal-ordered two-body (NO2B) approximation~\cite{PhysRevLett.126.022501} to the full 3NF has been widely used in nuclear {\it ab initio} calculations. In the NO2B approximation, the normal-ordered 3NF in the chosen reference is truncated at a two-body level. Namely, the approximation takes the zero-, one- and two-body parts of the 3NF in the normal-ordered form, but neglects the residual three-body term. The coupled cluster and importance-truncated no-core shell model calculations have shown that the contribution of the residual 3NF is not significant~\cite{PhysRevC.76.034302,PhysRevLett.109.052501}.

The total Hamiltonian~(\ref{Eq1}) is first transferred to the GHF basis, and then normal ordered with respect to the reference state which can take the GHF Slater determinant of the closed core. The NO2B Hamiltonian with neglecting the residual three-body term reads
\begin{equation}
\label{Eq5}
\begin{aligned}
\hat{H}=&\sum_{i=1}^A t_{ii}+\frac{1}{2}\sum\limits_{i,j=1}^A W^{\mathrm{NN}}_{ijij}+\frac{1}{6}\sum\limits_{i,j,k=1}^AW^{3\mathrm{N}}_{ijkijk}\\
&+\sum_{pq}(t_{pq}+\sum\limits_{i=1}^AW^{\mathrm{NN}}_{piqi}+\frac{1}{2}\sum\limits_{i,j=1}^AW^{3\mathrm{N}}_{pijqij}):\hat{a}_{p}^{\dagger} \hat{a}_{q}:\\
       &+\frac{1}{4} \sum_{pqrs} (W^{\mathrm{NN}}_{pqrs}+\sum_{i=1}^AW^{3\mathrm{N}}_{pqirsi}) :\hat{a}_{p}^{\dagger} \hat{a}_{q}^{\dagger} \hat{a}_{s} \hat{a}_{r}:,
\end{aligned}
\end{equation}
where $t_{ij}$ is the matrix element of the nucleon kinetic energy with a CoM-corrected mass $\mu=m/(1-\frac{1}{A})$, while $W^{\mathrm{NN}}_{pqrs}$ and $W^{3\mathrm{N}}_{pqirsi}$ are antisymmetric 2NF and 3NF matrix elements given in the GHF basis, respectively, which can be obtained by calculating overlaps between the GHF and HO basis states. $p,q,r,s$ and $i,j,k$ stand for generic states and hole states, respectively. $\hat{a}^\dagger$ and $\hat{a}$ represent the creation and annihilation operators, respectively. The colons indicate normal-ordering with respect to the reference state. Hamiltonian~(\ref{Eq5}) serves as the initial input of the many-body perturbation theory (MBPT)~\cite{HJORTHJENSEN1995125,CORAGGIO2009135}.


For the $A=16$ mirror partners $^{16}$F-$^{16}$N and $^{16}$C-$^{16}$Ne, we can choose the mirror-symmetric $^{14}\text{O}$ and $^{14}\text{C}$ as the closed cores. With the $^{14}\text{O}$ core, $^{16}$F has a valence proton and a valence neutron outside the closed core, while $^{16}$Ne has two valence protons. With the $^{14}\text{C}$ core, $^{16}$N has a valence proton and a valence neutron, while $^{16}$C has two valence neutrons. Therefore, no residual 3NF appears in the systems of two valence particles, though it may still appear in the calculations of the $\hat S$- and $\hat{Q}\text{-box}$ folded diagrams which are used to construct the valence-space effective Hamiltonian. However, the effect of the residual 3NF on the $\hat S$- and $\hat{Q}\text{-box}$ folded diagrams should be neglected. Note that the full 3NF was used in the HF and GHF calculations of the closed cores. The HF calculation starts within a HO basis at a frequency of $\hbar\omega =16$ MeV. This frequency was usually used in {\it ab initio} calculations with the EM1.8/2.0 interaction from light (He isotopes)  to heavy (Sn isotopes) nuclei~\cite{PhysRevLett.126.022501,PhysRevC.105.014302,PhysRevC.96.014303}. In the calculation, we set a basis truncation with $e=2n+l \leq e_{\mathrm{max}}=12$, and limit 3NF with $e_{1}+e_{2}+e_{3}\leq e_{3\mathrm{max}}=12$. After the HF calculation, we perform the GHF calculation as described above. For the $A=16$ mirror nuclei, the SM or GSM valence space can be chosen as $0p_{1/2}0d_{5/2}1s_{1/2}0d_{3/2}$. The $0p_{1/2}$ orbital should be well bound, while $0d_{5/2}$, $1s_{1/2}$ and $0d_{3/2}$ can be weakly bound or resonance. Therefore, we treat the $d_{5/2}$, $s_{1/2}$ and $d_{3/2}$ partial waves within the GHF basis which includes resonance and continuum, while other partial waves are treated within a discrete real-energy HF basis to reduce the computational task, which has been shown to be quite reasonable (see, e.g.,~\cite{PhysRevC.103.034305,PhysRevC.103.044319}. Finally, our GSM model space is defined as $\{\pi 0p_{1/2}, \pi 1s_{1/2}, \pi 0d_{5/2}, \pi 0d_{3/2}, \nu s_{1/2},\nu d_{5/2}, \nu d_{3/2}\}$ with the $^{14}\text{C}$ core, and $\{\pi s_{1/2},\pi d_{5/2}, \pi d_{3/2}, \nu 0p_{1/2}, \nu 1s_{1/2}, \nu 0d_{5/2}, \nu 0d_{3/2}\}$ with the $^{14}\text{O}$ core. 

We use the many-body perturbation theory (MBPT) to derive the effective Hamiltonian in the valence space, i.e., one-body $\hat S$-box and two-body $\hat{Q}\text{-box}$ folded diagrams~\cite{CORAGGIO200543,TAKAYANAGI201161,PhysRevC.89.024313}. The $\hat{S}$ box is by definition the one-body part of the $\hat Q$ box. The MBPT calculation has been extended to the complex-$k$ Berggren basis~\cite{PhysRevC.80.051301,SUN2017227,HU2020135206}. Due to the large number of discretized continuum states, the inclusion of third-order $\hat{Q}\text{-box}$ diagrams is computationally very costly and often renders MBPT calculations impossible. Therefore, we calculate the $\hat{S}\text{-box}$ folded diagrams up to the third order, while $\hat{Q}\text{-box}$ folded diagrams up to the second order, which has been shown to offer a good approximation~\cite{HU2020135206}. For nuclear excitation spectra, the effect from the neglected third-order $\hat Q$-box folded diagrams is rather small compared with the effect on the total binding energy~\cite{HU2020135206}. 

After the $\hat S$-box and $\hat Q$-box calculations, we obtain the GSM effective Hamiltonian in the chosen complex-energy Berggren valence space, 
\begin{equation}
\begin{aligned}
\hat{H}_{\text{eff}}=&\sum_{pq}\epsilon_{pq}\hat{a}_{p}^{\dagger} \hat{a}_{q}
&+\frac{1}{4} \sum_{pqrs} V_{pqrs}^{\text{eff}} \hat{a}_{p}^{\dagger} \hat{a}_{q}^{\dagger} \hat{a}_{s} \hat{a}_{r},
\end{aligned}
\end{equation}
where $p,q,r,s$ indicate valence particles in the valence space. $\epsilon_{pq}$ stands for valence-space effective single-particle Hamiltonian matrix elements obtained by $\hat S$-box with the GHF Hamiltonian~(\ref{Eq3}), which correspond
to valence single-particle energies 
but containing some nonzero (but small) off-diagonal matrix elements due to the coupling in the same partial wave. $V_{pqrs}^\text{eff}$ is the effective interaction matrix elements (including the NO2B 3NF) derived by the $\hat Q$ box. The complex-symmetric GSM effective Hamiltonian is diagonalized in the valence space using the Jacobi-Davidson method in the $m$-scheme~\cite{MICHEL2020106978}. 


\section{\label{sec:level}Calculations and discussions}

\begin{figure}[b]
\includegraphics[width=0.48\textwidth]{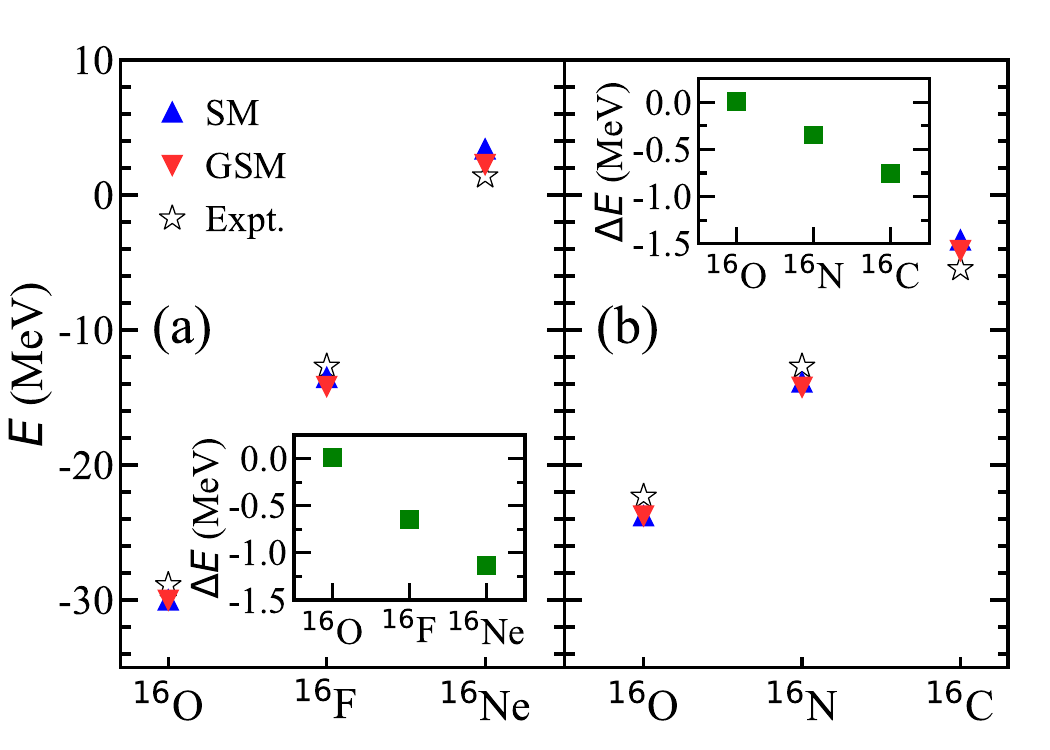}
\caption{\label{16be} Calculated and experimental ~\cite{ensdf} g.s. energies of $^{16}$O, $^{16}$F and $^{16}$Ne with respect to the $^{14}$O core (a), and $^{16}$O, $^{16}$N and $^{16}$C with respect to the $^{14}$C core (b). The 2NF plus 3NF interaction EM1.8/2.0 is used. Standard real-energy SM calculations with the same EM1.8/2.0 have also been performed to see the continuum effect defined by $\Delta E=E_\text{GSM}-E_\text{SM}$ shown in the insets.} 
\end{figure}

Figure~\ref{16be} shows calculated and experimental~\cite{ensdf} g.s. energies for $^{16}$O, $^{16}$F and $^{16}$Ne with respect to the closed $^{14}$O core (left panel) and for $^{16}$O, $^{16}$N and $^{16}$C with respect to the closed $^{14}$C core (right panel). We see that the GSM calculations well reproduce the experimental energies of the $A=16$ mirror nuclei. To see the continuum effect, we have also performed standard SM calculations in the real-energy HF basis with the same EM1.8/2.0. The real-energy calculation means that no continuum coupling is included. For the energies, the standard SM can also give quite good results. By comparing the energy difference between the GSM and SM calculations, we can see the continuum effect as shown in the insets of Fig.~\ref{16be}. The continuum coupling lowers the energies of $^{16}$F, $^{16}$Ne, $^{16}$N and $^{16}$C by $\approx$ 0.4-1.1 MeV, while the energy of the deeply bound $^{16}$O is nearly unchanged whether the continuum coupling is included. 


The SM (GSM) gives the energy of an open-shell valence-particle system outside the closed core. This energy is relative to the core. For the total energy of an open-shell nucleus, one needs to know the core energy. SM (GSM) itself is not able to produce the core energy, while the HF (GHF) calculation based on a realistic interaction is not sufficient to describe the core. Higher-order correlations beyond the HF approximation should be considered. This can be done using the many-body Rayleigh-Schrödinger perturbation theory (RSPT)~\cite{PhysRevC.94.014303}. As commented in Refs.~\cite{PhysRevC.94.014303}, the RSPT corrections up to the third order can well describe the energy of a closed nucleus. With the same EM1.8/2.0 interaction, the RSPT calculation based on the HF approximation gives the g.s. energies of $-96.19$ MeV and $-102.36$ MeV for the $^{14}$O and $^{14}$C cores, respectively, almost the same as the values of $-96.16$ MeV and $-102.38$ MeV calculated by the {\it{ab initio}} in-medium similarity renormalization group (IMSRG) with the same interaction. The calculated energies are also in reasonable agreements with the data of $-98.73$ MeV and $-105.28$ MeV~\cite{ensdf} for $^{14}$O and $^{14}$C, respectively. For $^{16}$O as shown in Fig.~\ref{16be}, the left panel gives the energy of the two valence neutrons outside the closed $^{14}$O core, while the right panel shows the energy of the two valence protons outside the closed $^{14}$C core. Added to the core energies calculated by RSPT, we obtain the $^{16}$O total g.s. energy: $-126.20$ MeV with the $^{14}$O core or $-126.14$ MeV with the $^{14}$C core. The two calculated energies for $^{16}$O are almost the same, and also agree well with the experimental g.s. energy of $-127.62$ MeV~\cite{ensdf}.

\begin{figure}[t]
\includegraphics[width=0.5\textwidth]{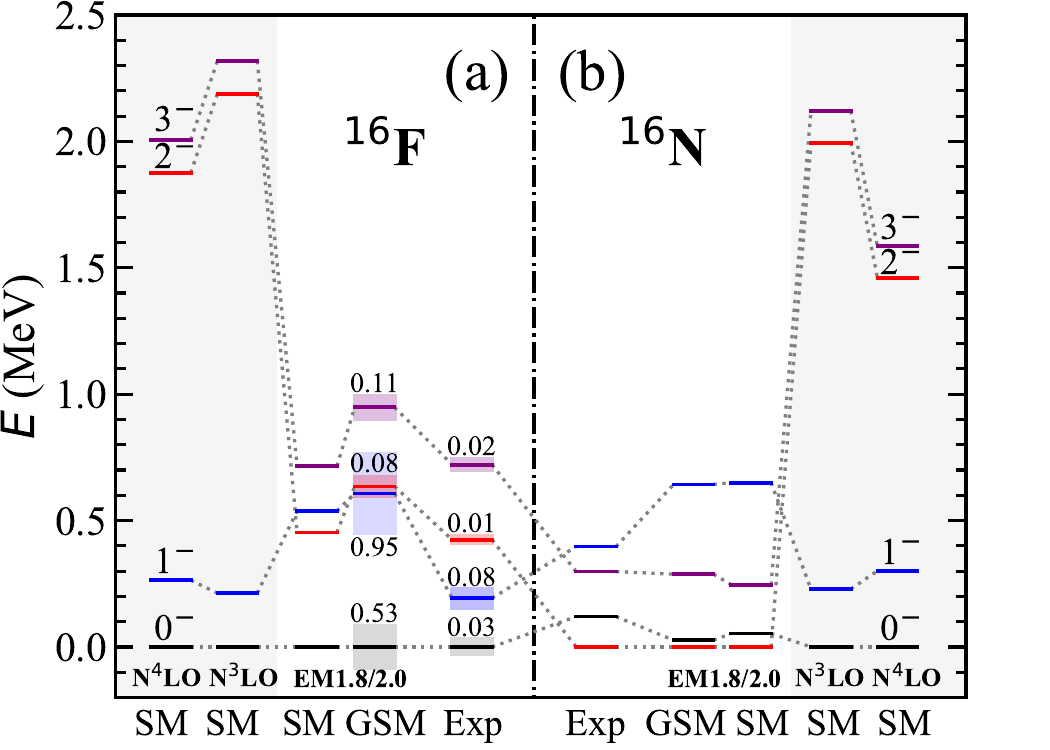}
\caption{\label{fn16} Low excitation spectra of the mirror partners $^{16}\text{F}$ and $^{16}\text{N}$. N$^{3}$LO and N$^{4}$LO indicate only bare 2NF used in the calculations. EM1.8/2.0 includes 3NF. SM and GSM mean the continuum coupling excluded and included in calculations, respectively. The shadow indicates a resonance with the resonance width (in MeV) given by the number above (below) the level. Experimental data are taken from \cite{PhysRevC.90.014307,ensdf}.}
\end{figure}

Experiment data ~\cite{PhysRevC.90.014307,ensdf} give very different excitation spectra with different g.s. levels and different level orderings between the pair of the mirror partners $^{16}$F and $^{16}$N, and the pronounced TES's are seen, as shown in Fig.~\ref{fn16}.
With the continuum coupling included, the recent coupled-channel GSM calculation with phenomenological potentials can produce the correct ordering of excitation levels in $^{16}$F and $^{16}$N in the presence of {\it ad hoc} corrective factors~\cite{PhysRevC.106.L011301}. Here, we employ the self-consistent {\it ab initio} GSM with the chiral 3NF included to investigate the 3NF and continuum effects on the spectra of the mirror partners $^{16}$F and $^{16}$N, including the observed TES's. Figure~\ref{fn16} shows the low-lying excitation spectra calculated by different models or interactions. Calculations start with the real-energy standard SM and only 2NF considered (bare chiral N$^{3}$LO~\cite{PhysRevC.68.041001} and N$^{4}$LO~\cite{PhysRevC.96.024004} at $\Lambda =500$ MeV). These calculations 
mean that the continuum and 3NF effects are not included. For the $^{16}$F spectrum, the level orderings obtained with N$^{3}$LO and N$^{4}$LO are in agreements with the experimental spectrum~\cite{PhysRevC.90.014307,ensdf}, but the calculated $2^-$ and $3^-$ levels are too high compared with the data, see Fig.~\ref{fn16}. Similarly, SM calculations with N$^{3}$LO and N$^{4}$LO also give too high $2^-$ and $3^-$ levels in the mirror partner $^{16}$N, as shown in Fig.~\ref{fn16}. In the experiment, $^{16}$F has an unbound resonant $0^-$ g.s., while the mirror partner $^{16}$N has a bound $2^-$ ground state. This phenomenon that the mirror partners have different ground states in configurations cannot be reproduced in the 2NF-only SM calculations which instead give the same $0^-$ g.s. for the pair of mirror nuclei. For the mirror partner $^{16}$N, the experiment gives different level ordering from that in $^{16}$F, while the SM calculations with 2NF-only give the same level ordering for this pair of mirror partners. It then shows the inability of SM calculations to reproduce the experimental data using 2NF only. Since the contribution of the continuum coupling is much smaller than the theoretical and experimental differences in the energy spectra, we do not perform GSM calculations with 2NF only. 

With the 3NF included (e.g., using the EM1.8/2.0 interaction), the agreements with experimental spectra are significantly improved in both SM and GSM calculations for the mirror partners $^{16}$F and $^{16}$N, as shown in Fig.~\ref{fn16}. The correct level ordering in the $^{16}$N spectrum is reproduced in both SM and GSM calculations when 3NF is included. In $^{16}$F, however, the SM calculation without the continuum effect included cannot give the correct order between the $1^-$ and $2^-$ levels compared with data, while the GSM calculation with both 3NF and continuum coupling included provides a correct order of the two levels. The GSM calculation shows that the $1^-$ excited state in $^{16}$F is mainly composed of the $\pi s_{1/2}\otimes\nu 0p_{1/2}$ configuration, whereas the $2^-$ excited state primarily consists of $\pi d_{5/2}\otimes\nu 0p_{1/2}$. Since there is no centrifugal barrier with $l=0$, the wave function of the $s$-wave is more spread in space, resulting in a stronger coupling to the continuum, and thus a lower energy for the $1^-$ state. This corrects the level order between the $1^-$ and $2^-$ states. From the experimental and calculated resonance widths of the unbound $^{16}\text{F}$ states, indeed, the states containing a significant component of the $s$ partial wave, e.g., the $0^-$ and $1^-$ levels, have broader resonance widths. Therefore, to describe the nature of weakly bound and unbound nuclei, a rigorous treatment of the asymptotic behavior of the single-particle wave functions and their coupling to the scattering continuum is necessary. Nevertheless, for $^{16}\text{F}$ and $^{16}\text{N}$, the 3NF is the first important factor to improve the spectrum description, as shown in Fig.~\ref{fn16}.

\begin{figure}[b]
\includegraphics[width=0.48\textwidth]{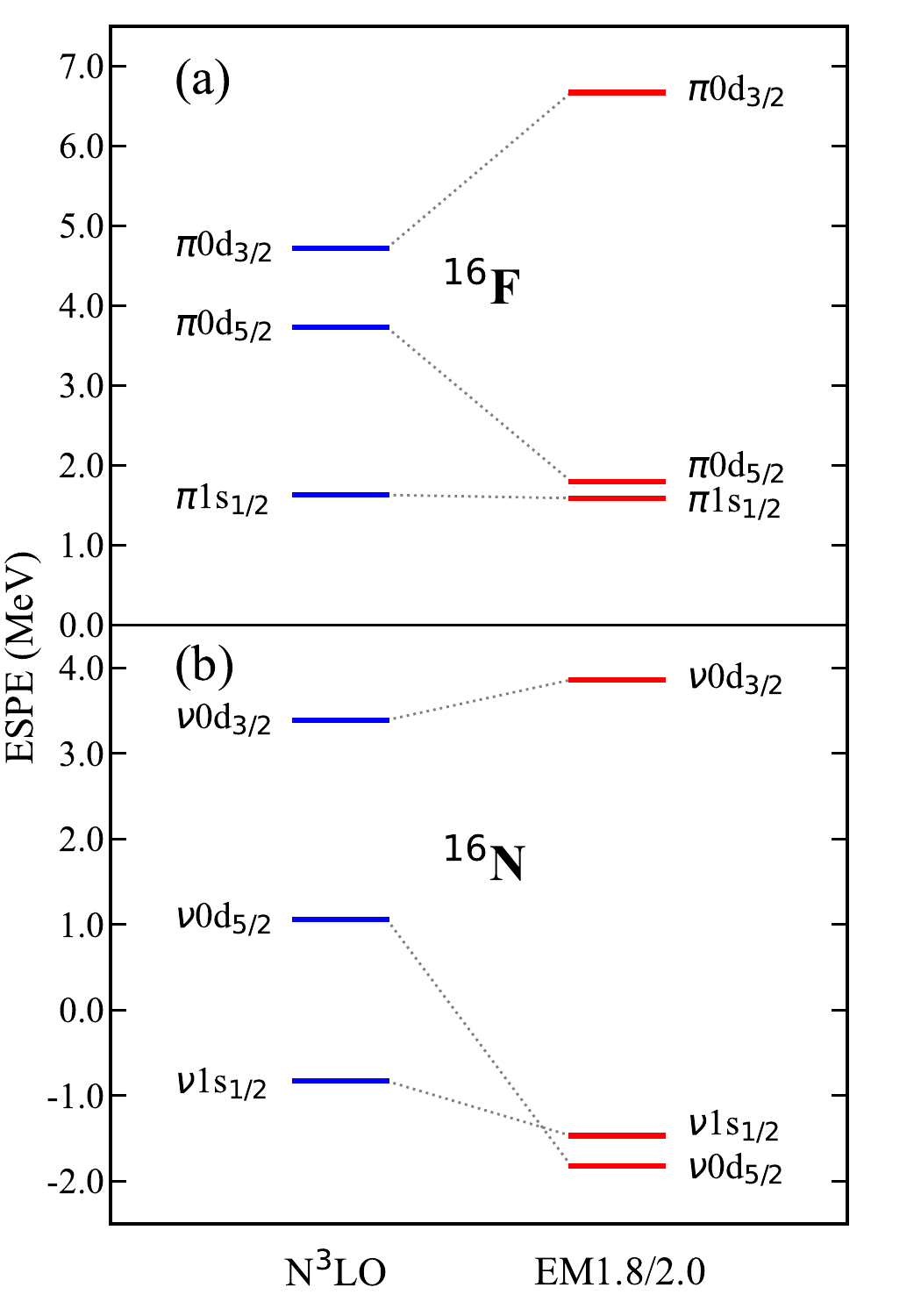}
\caption{\label{espe} Effective single-particle energies (ESPEs) of the {\it{sd}} proton ($\pi$) orbitals in $^{16}\text{F}$ (a) and {\it{sd}} neutron ($\nu$) orbitals in $^{16}\text{N}$ (b), calculated by SM with the 2NF N$^3$LO and the 2NF plus 3NF EM1.8/2.0 interactions.} 
\end{figure}

We see in Fig.~\ref{fn16} that the inclusion of 3NF causes a large energy drop in the $2^-$ and $3^-$ states of the mirror partners $^{16}$F and $^{16}$N. This can be understood by calculating effective single-particle energies (ESPEs)~\cite{PhysRevC.100.034324} with the same EM1.8/2.0, as shown in Fig.~\ref{espe}. $^{16}$F has a dominant configuration of $\pi 1s_{1/2}\otimes \nu 0p_{1/2}$ for the $0^-$ g.s. and $1^-$ excited state, and a primary configuration of $\pi 0d_{5/2}\otimes \nu 0p_{1/2}$ for the $2^-$ and $3^-$ excited states, while $^{16}$N has a dominant configuration of $\nu 1s_{1/2}\otimes \pi 0p_{1/2}$ for the $0^-$ and $1^-$ states, and a main configuration of $\nu 0d_{5/2}\otimes \pi 0p_{1/2}$ for the $2^-$ and $3^-$ states. As shown in Fig.~\ref{espe}, the 3NF causes a large energy drop in both proton and neutron $0d_{5/2}$ orbitals, which leads to large energy drops in the $2^-$ and $3^-$ states. In $^{16}$N, the neutron $0d_{5/2}$ orbital is even lower than $1s_{1/2}$ when 3NF is included, which results in a $2^-$ g.s. instead a $0^-$ g.s. in $^{16}$N.

\begin{figure}[t]
\includegraphics[width=0.48\textwidth]{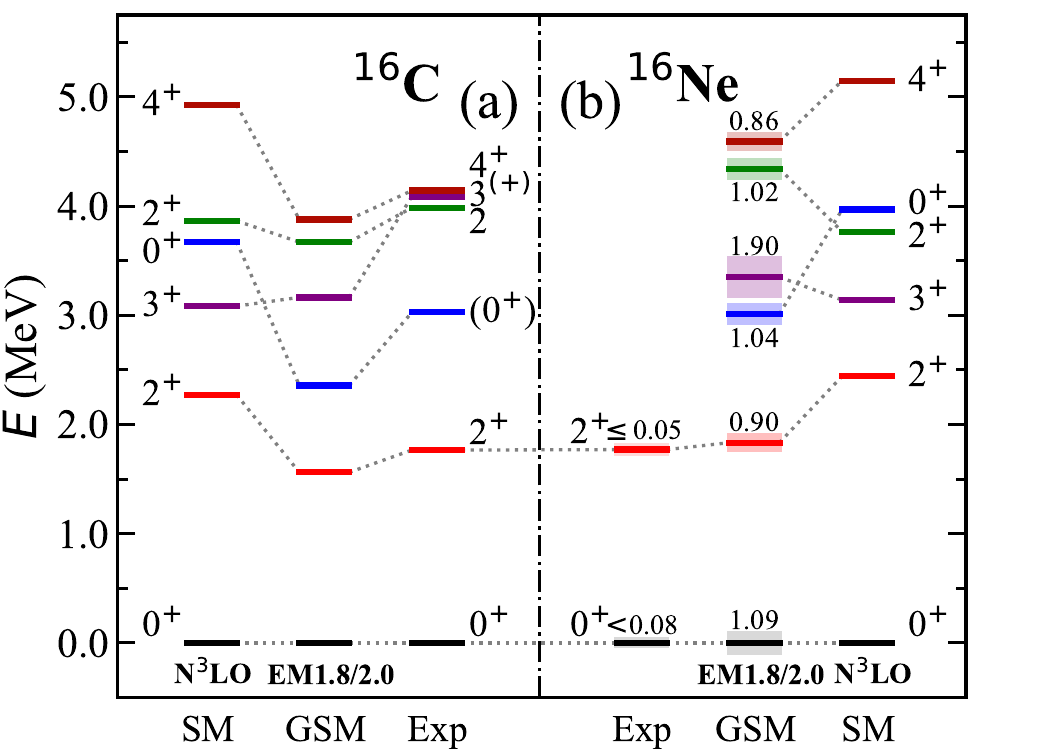}
\caption{\label{cne16}  Similar to Fig.~\ref{fn16}, but for the even-even mirror partners $^{16}\text{C}$ and $^{16}\text{Ne}$. Experimental data are taken from \cite{PhysRevC.86.044329,PhysRevLett.113.232501,ensdf}.} 
\end{figure}

$^{16}$C and $^{16}$Ne form a pair of even-even $A=16$ mirror partners. The works~\cite{PhysRevLett.88.042502,PhysRevC.91.024325} using a three-body model had investigated the $0^+$ g.s. and the first $2^+$ excited state, commenting that significant structure differences exist between the mirror isobaric analog states (IAS) in the mirror partners $^{16}$C and $^{16}$Ne, and was suggested as an additional dynamic TES mechanism. With both 3NF and the continuum coupling included, we have performed the self-consistent {\it ab initio} many-body GSM calculations with the same chiral EM1.8/2.0 interaction, as shown in Fig.~\ref{cne16}. We see that the 3NF and continuum coupling are important to reproduce the experimental spectra including the level order and resonance nature. 

\begin{figure}[b]
\includegraphics[width=0.48\textwidth]{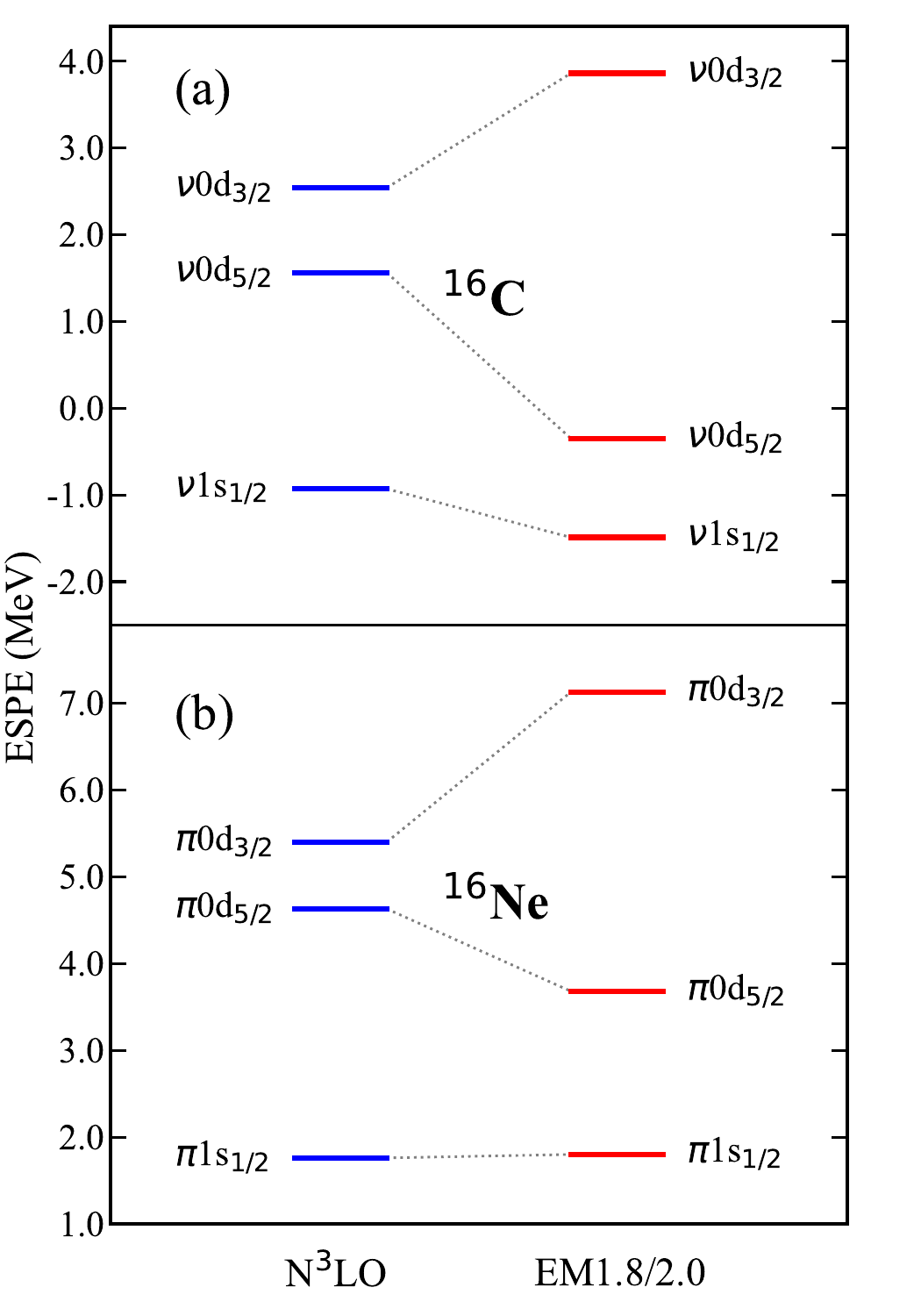}
\caption{\label{espeCNe} Similar to Fig.~\ref{espe}, but for {\it{sd}} neutron orbitals in $^{16}\text{C}$ (a) and {\it{sd}} proton orbitals in $^{16}\text{Ne}$ (b).} 
\end{figure}

\begin{figure}[]
\includegraphics[width=0.48\textwidth]{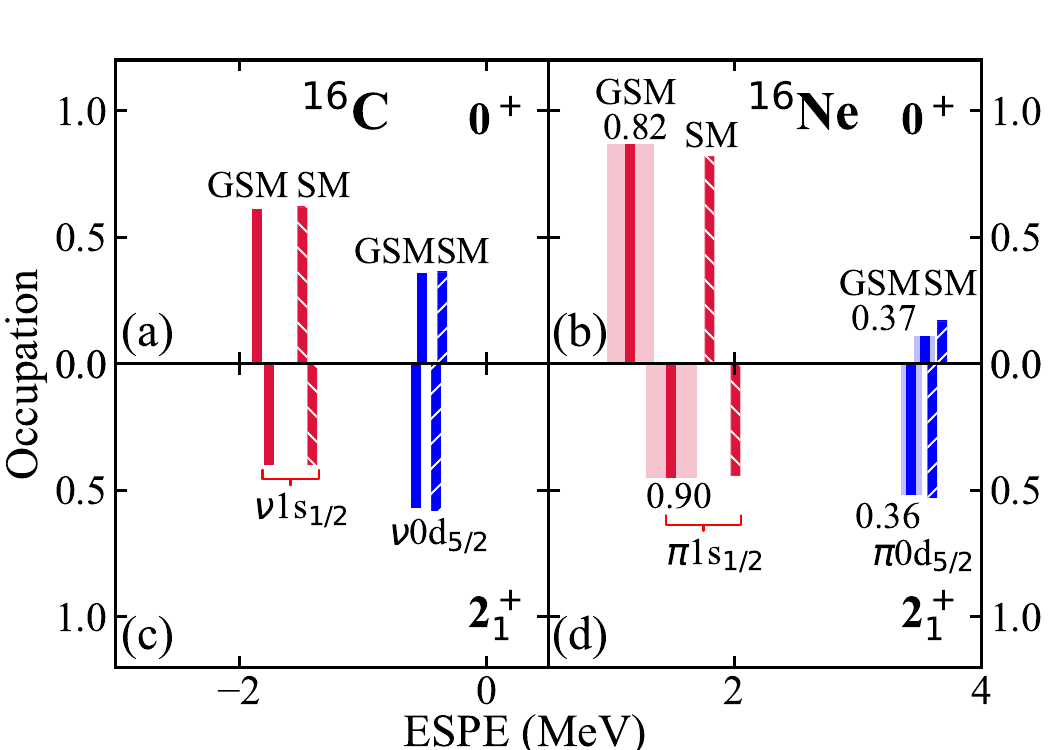}
\caption{\label{occupation}Neutron ($\nu$) or proton ($\pi$) $1s_{1/2}$ and $0d_{5/2}$ ESPEs, and nucleon occupation probabilities on $s_{1/2}$ and $d_{5/2}$ partial waves in the $0^+$ g.s. [upper (a), (b)] and $2^+_1$ excited state [lower (c), (d)] for $^{16}\text{C}$ [left (a), (c)] and $^{16}\text{Ne}$ [right (b), (d)]. In the GSM calculation, $\pi 1s_{1/2}$ and $\pi 0d_{5/2}$ are resonances, indicated by shadowing with the resonance width given above (below) the bar. The EM1.8/2.0 interaction is used.} 
\end{figure}

Figure~\ref{espeCNe} shows the ESPEs in $^{16}$C and $^{16}$Ne. We see that the energy drop also appears in the neutron and proton $0d_{5/2}$ orbitals when 3NF is included in the calculations. However, the drop is not as large as that seen in the odd-odd $^{16}$F and $^{16}$N mirror nuclei. The energy drop can also be found in the calculated spectra, as shown in Fig.~\ref{cne16}. The mechanism of the energy drop is similar to that occurring in the odd-odd $^{16}$F and $^{16}$N. For example, the first $2^+$ excited state in $^{16}$C and $^{16}$Ne contains a large $0d_{5/2}$ component (see Fig.~\ref{occupation}), therefore we see the energy drop in the $2^+_1$ levels when 3NF is included in the calculations. 

Figure~\ref{occupation} shows the dominant $0d_{5/2}$ and $1s_{1/2}$ configurations in the g.s. and first $2^+$ excited states of $^{16}$C and $^{16}$Ne. The continuum coupling lowers the energies of the $1s_{1/2}$ and $0d_{5/2}$ orbitals, with the $1s_{1/2}$ orbital being lowered more significantly due to the strong coupling to the continuum in the $l=0$ partial wave. The ground states have more $1s_{1/2}$ component than the $2^+$ excited states, therefore the g.s. energy drop due to the continuum coupling is more significant than the $2^+$ state, which in turn raises the $2^+$ excitation energy which is relative to the g.s. energy. Therefore, the actual $2^+$ energy drop due to 3NF is larger than that shown in Fig.~\ref{cne16} which plots the combined effect from 3NF and the continuum coupling. The 3NF-caused energy drop in even-even $^{16}$C and $^{16}$Ne is not as pronounced as in odd-odd $^{16}$F and $^{16}$N. In addition to the less pronounced $0d_{5/2}$ energy drop in $^{16}$C and $^{16}$Ne, a less pronounced $d_{5/2}$ component is another reason for the less pronounced energy drops in even-even $^{16}$C and $^{16}$Ne. The states in $^{16}$C and $^{16}$Ne do not purely occupy the $d_{5/2}$ partial wave, but also contain significantly other partial waves, e.g., the $s_{1/2}$ partial wave, as shown Fig.~\ref{occupation}.

We predict a resonance excitation spectrum for $^{16}$Ne with a similar level scheme to its mirror partner $^{16}$C, but significant TES's are seen in the $0_2^+$, $2_2^+$ and $4^+$ states. Calculated resonance widths for the observed $0^+$ g.s. and $2^+$ excited state are broader than data in the unbound resonant nucleus $^{16}$Ne. The resonance width is sensitive to the separation energy of particle emissions~\cite{PhysRevC.103.044319}, and the present GSM calculation of the g.s. energy is more unbound than the experimental datum, therefore broader widths have been obtained. However, the present predictions of the excitation energies and resonance widths in $^{16}$Ne are similar to those of the GSM calculations using phenomenological potentials~\cite{PhysRevC.103.044319}.

In the low-lying states of the odd-odd $^{16}$F and $^{16}$N, due to the magic shell gap above $0p_{1/2}$, the odd neutron in $^{16}$F or the odd proton in $^{16}$N mainly occupies the $0p_{1/2}$ orbital, while the odd proton in $^{16}$F or the odd neutron in $^{16}$N stays in the {\it sd} shell with our defined model space, which leads to negative-parity low excitation levels as shown in Fig.~\ref{fn16}. For the even-even $^{16}$C and $^{16}$Ne, the pair of valence neutrons or valence protons stay in the {\it sd} shell defined by our model space. Therefore, lowly excited states in $^{16}$C and $^{16}$Ne should have a positive parity. Negative-parity levels should have higher excitation energies, with valence particle(s) being excited to the higher {\it pf} shell above the {\it sd} shell or hole excitation in the {\it p} shell of the core. 

In Refs.~\cite{PhysRevLett.88.042502,PhysRevC.91.024325}, the mirror symmetry breaking and TES in the lowest $0^+$ and $2^+$ states of the mirror partners $^{16}$Ne and $^{16}$C were studied using a three-body model with phenomenological interactions. Here, we analyze the mirror symmetry breaking using the wave functions obtained in our {\it ab initio} calculations. The low-lying states are governed by the $s_{1/2}$ and $d_{5/2}$ partial waves, as shown in Fig.~\ref{occupation} for the $0^+_1$ and $2^+_1$ states in $^{16}$Ne and $^{16}$C. The neutron $1s_{1/2}$ and $0d_{5/2}$ orbitals are weakly bound, while the proton $1s_{1/2}$ and $0d_{5/2}$ orbitals are unbound resonances (Note that the real-energy standard SM calculation cannot provide the description of the resonance nature). We see that the continuum effect on the $1s_{1/2}$ orbital is more pronounced with a significant energy shift between the GSM and SM ESPE's, as shown in Fig.~\ref{occupation}. This is because no centrifugal barrier exists in the $s$-wave, which leads to a stronger coupling to the continuum. We also see that, in the $0^+$ g.s., the occupation probability on the $s_{1/2}$ partial wave in $^{16}$Ne is significantly larger than that in the mirror partner $^{16}$C, while this mirror asymmetry does not appear to be obvious in the $2^+$ IAS (shown in the lower panel). This is in consistence with the statement given in Refs~\cite{PhysRevLett.88.042502,PhysRevC.91.024325}. The unbound $^{16}$Ne $0^+$ g.s. contains more $s_{1/2}$ component, and thus has a stronger continuum coupling effect, which leads to a more energy drop than in other states. Consequently, this raises the $2^+$ excitation energy of the GSM calculation, which may provide an explanation for the recent experimental observation that the 2$^+$ state in $^{18}$Mg is higher than that in $^{18}$C~\cite{PhysRevLett.127.262502}.

Finally, we test the stability or convergence of the calculations in which we have to take some approximations or truncations due to limitations of computing power. The calculations start with the HF (GHF) approximation using the same realistic interaction, which provides the basis for many-body calculations. When the discretized continuum partial waves are included in the valence space, it is difficult to include the third-order $\hat Q$-box perturbation diagrams due to the huge model dimension. In our previous works~\cite{HU2020135206,Wu_2019}, however, it has been shown that the effect from the third-order $\hat Q$-box folded diagrams is rather small on the spectrum. The HO $\hbar\omega$ value which was taken at the beginning to generate interaction matrix elements is another factor that should be checked to see whether the calculation is heavily dependent on the $\hbar\omega$ value used. In Ref.~\cite{PhysRevLett.126.022501}, the EM1.8/2.0 interaction with $\hbar\omega=16$ MeV was used for the {\it ab initio} calculations of nuclei from He to Fe isotopes, showing good agreements with data. The phenomenological $\hbar\omega=45A^{-1/3}-25A^{-2/3}$ gives $\hbar\omega\approx 14$ MeV for $A=16$ mass. The generation of the interaction matrix elements in the complex-$k$ space is a huge task due to the huge dimension with the large number of discretized continuum states. However, we have tested the $\hbar\omega$ dependence of calculations in the real-energy space. Using $\hbar\omega=14$ MeV, 16 MeV and 20 MeV with the same EM1.8/2.0 interaction, we find that obtained HF single-particle energies remain almost the same, and the RSPT calculations with the inclusion of the third-order correction give the total energy of $-102.12$ MeV, $-102.36$ MeV and $-102.39$ MeV for the $^{14}$C core, respectively. Therefore, the dependence on the $\hbar\omega$ value should not be serious in the calculations. With the EM1.8/2.0 interaction at $\hbar\omega=16$ MeV, we have tried a larger 3NF truncation with $e_{\text{3max}}=14$ and a denser contour discretization with 40 discretizing points for each continuum partial wave, which reaches the maximum limit of our computer resources. Obtained GSM results are almost the same as those obtained using $e_{\text{3max}}=12$ and 35 discretizing points. The maximum difference in the g.s. energies is 14 keV occurring in $^{16}$N, while it is only 4 keV in the excitation energies of spectra occurring in the $1^-$ level of $^{16}$N. 

\section{\label{sec:level4}Summary}

Nuclei around driplines exhibit unique features with resonance and strong coupling to the continuum. Using the {\it ab initio} Gamow shell model with chiral three-nucleon force included, we have investigated the mirror asymmetry in $A=16$ mirror pairs. This model provides a comprehensive and rigorous {\it ab initio} description of many-body correlations in the presence of the continuum coupling.
As compared to previously used Gamow shell models, the present calculations make use of the Gamow Hartree-Fock basis based on the initial chiral two- and three-nucleon forces. With both the three-nucleon force and the continuum coupling included, the calculations can well reproduce the experimental level inversions in the $^{16}$F-$^{16}$N mirror pair. Comparisons with other models demonstrate that both the three-nucleon force and the continuum coupling should be included in the {\it ab initio} calculations of weakly bound and unbound open quantum nuclear systems. Our calculations have also provided a rigorous test of the EM1.8/2.0 nuclear force and the {\it ab initio} method used.
The calculations reasonably reproduce the experimental low-lying excited states of $^{16}$C except the $3^+$ level, and predict the low-lying spectrum of its mirror partner $^{16}$Ne which is resonance. We calculated effective single-particle energies and analyzed occupation probabilities on partial waves, which reveals that a Thomas-Ehrmann shift develops in the $0^+$ ground-state configurations of $^{16}$C-$^{16}$Ne mirror pairs. The study of the mirror asymmetry in the $^{16}$C-$^{16}$N mirror pair can also provide further insight into the origin of the relatively high 2$^{+}$ excitation energy in $^{18}$Mg.


\section*{\label{sec:level7} Acknowledgments}

Valuable discussions with Simin Wang are gratefully acknowledged.
This work has been supported by 
the National Natural Science Foundation of China under Grants No. 12335007, 11835001, 11921006, 12035001, 12205340 and 12135017; the Gansu Natural Science Foundation under Grant No. 22JR5RA123; the State Key Laboratory of Nuclear Physics and Technology, Peking University under Grant No. NPT2020ZZ01. We acknowledge the High-Performance Computing Platform of Peking University for providing computational resources.

\bibliography{article}

\end{document}